# Introduction of Zr in nanometric periodic Mg/Co multilayers

K. Le Guen[1*], M.-H. Hu[1], J.-M. André[1], P. Jonnard[1], S. K. Zhou[2], H. Ch. Li[2],

J. T. Zhu[2], Z. S. Wang[2], N. Mahne[3], A. Giglia[3], S. Nannarone[3]

[1]Laboratoire de Chimie Physique – Matière Rayonnement, UPMC Univ Paris 06, CNRS UMR 7614
11 rue Pierre et Marie Curie, F-75231 Paris cedex 05, France

[2]Institute of Precision Optical Engineering. Department of Physics, Tongji University,
Shanghai 200092, P.R. China

[3]Istituto Officina dei Materiali IOM-CNR Laboratorio TASC
S.S.14. km 163.5 in Area Science Park, I-34012 Trieste, Italy

**Date: 16/07/2010**

[*]Corresponding author:
Dr Karine Le Guen, Laboratoire de Chimie Physique – Matière et Rayonnement, 11 rue Pierre et Marie Curie, F-75231 Paris Cedex 05, France; tel: 33 1 44 27 66 08; fax: 33 1 44 27 62 26
karine.le_guen@upmc.fr

**Abstract**

We study the introduction of a third material, namely Zr, within a nanometric periodic Mg/Co structure designed to work as optical component in the extreme UV (EUV) spectral range. Mg/Co, Mg/Zr/Co, Mg/Co/Zr and Mg/Zr/Co/Zr multilayers are designed, then characterized in terms of structural quality and optical performances through X-ray and EUV reflectometry measurements respectively. For the Mg/Co/Zr structure, the reflectance value is equal to 50% at 25.1 nm and 45° of grazing incidence and reaches 51.3% upon annealing at 200°C. Measured EUV reflectivity values of tri-layered systems are discussed in terms of material order within a period and compared to the predictions of the theoretical model of Larruquert. Possible applications are pointed out.

# 1. Introduction

We are interested in the design, the development and the characterization of multilayered mirrors exhibiting high reflectance values in the extreme ultraviolet (EUV) range [1-4]. To that purpose, we explore different ways as for instance the introduction of a third material within a bilayered structure and the effect of annealing. Multilayers made of more than two materials have already proven to be efficient optical systems [5-10].

Such multilayered mirrors can be used as optical components in various applications owing to the high peak reflectivity at the wavelength of interest which is enhanced with respect to the neighbouring wavelengths although the spectral purity is not assured. They can be used as monochromators at 45° in synchrotron centres, in telescopes for imaging purposes even if telescopes are generally operated at normal incidence.

We have recently shown that Co/Mg multilayers exhibit a reflectance value equal to 42.6% at 25.1 nm [3]. To enhance the performance of this system, in the following we present the comparative analysis of the optical performances in the EUV spectral range of bi-, tri- and quadri-layered structures based on Mg, Co and Zr. X-ray and EUV reflectivity measurements are performed to access the structural parameters of the stack and estimate the EUV optical performances respectively. Measured EUV reflectivity values of tri- and quadri-layered systems are discussed in terms of material order within the structure on the basis of the theoretical model of Larruquert [8, 11]. The effect of annealing up to 200°C on the reflectance of the Mg/Co and Mg/Co/Zr multilayers is also studied.

## 2. Experiment

*2.1 Multilayer deposition*

The studied periodic multilayers were prepared using a calibrated ultra-high vacuum direct current magnetron sputtering system (JGP560C6. SKY Inc. China) with targets of Mg (purity 99.98%), Co (purity 99.95%), and Zr (purity 99.5%) in Ar gas (99.999%). The targets are 100 mm in diameter. The base pressure was $5.10^{-5}$ Pa and the working pressure was $1.10^{-1}$ Pa of Ar gas. The power applied on the Co, Mg and Zr targets was set to 20, 15 and 20 W respectively. Multilayers were deposited onto 30 mm x 40 mm ultra-smooth polished Si substrates with *rms* surface roughness of 0.3 nm. Each sample is made of 30 periods. The first layer on the substrate is the first layer given in the sample name. A 3.50 nm-thick $B_4C$ capping layer is deposited at the surface of each sample to prevent oxidation.

Materials and layer thicknesses are optimized to get the highest reflectance at 45° of incidence. At 45°, the Bragg peak is very close to the Mg L absorption edge (24.9 nm for $2p_{1/2}$ and 25.0 nm for $2p_{3/2}$) and consequently its shape is asymmetrical. That is why we consider, in addition to the 45° value, the 50° grazing angle in order to move the Bragg peak away from the Mg L edge. The spectra measured at 50° will be used for fitting purpose while those measured at 45° will allow us to estimate the optical performance of the multilayer at the application angle and wavelength. The structure of each designed sample and the IMD simulated reflectivity of the "ideal" stack (neither roughness nor interfacial compound) are detailed in Table 1 [12].

.

Table 1: structure and simulated reflectance of the "aimed" multilayers. IMD reflectivity simulations are performed assuming an "ideal" structure (neither roughness nor interfacial compound).

| Multilayer | Period d (nm) | $d_{Mg}$ (nm) | $d_{Co}$ (nm) | $d_{Zr}$ (nm) | Simulated Reflectivity $\theta_{grazing} = 45°$ | Simulated reflectivity $\theta_{grazing} = 50°$ |
|---|---|---|---|---|---|---|
| Mg/Co | 17.00 | 14.45 | 2.55 | - | 56.7% @ 25.2 nm | 54.0% @ 26.4 nm |
| Mg/Zr/Co | 17.20 | 13.20 | 2.50 | 1.50 | 54.6% @ 25.2 nm | 52.4% @ 26.5 nm |
| Mg/Co/Zr | 17.20 | 13.20 | 2.50 | 1.50 | 62.7% @ 25.2 nm | 58.3% @ 26.4 nm |
| Mg/Zr/Co/Zr | 17.00 | 12.00 | 2.00 | 1.50 | 47.1% @ 25.2 nm | 53.1% @ 26.0 nm |

From the results of the reflectivity simulations, some general trends can be drawn:

- at 45° of grazing incidence, whatever the structure is, the reflectivity is maximum at 25.2 nm owing to the Mg L discontinuity;
- with respect to the bi-layered structure, the addition of a third material (Zr) to form tri- and quadri-layers does not systematically lead to a higher reflectance;
- among the tri-layered structures, the order of the material sequence seems to play an important role: here, the Mg/Co/Zr system leads to a higher reflectance value than that of Mg/Zr/Co.

*2.2 X-ray reflectivity at 0.154 nm*

The quality of the multilayers was controlled through x-ray reflectometry using the Cu K$\alpha$ emission line (0.154 nm or 8048 eV). Measurements are made using a grazing incidence X-ray reflectometer (D1 system. Bede Ltd) operated in the $\theta$-2$\theta$ mode. The angular resolution is 5/1000°. Bragg law corrected for refraction was used to determine the multilayer period. The fit of the XRR curves performed with IMD was used to determine the thickness and roughness of the different layers in each structure and also to estimate their density. The roughness deduced from the fit is an overall roughness including the contributions from both geometrical roughness and interdiffusion.

*2.3 EUV reflectivity*

The measurement of the reflectivity curves in the EUV domain is performed on the BEAR beamline [13] at the Elettra synchrotron centre using *s*-polarized light. The photon energy is carefully calibrated using the Pt $4f_{7/2}$ feature and the Si L edge. The goniometer angular resolution is 1/100°. Impinging and reflected photon intensities are measured using a photodiode. Incident intensities are monitored using an Au mesh inserted in the beam path whose drain current is used for normalization. The overall accuracy on the absolute reflectivity values is estimated to be about 1%. A short (22.5-27.6 nm at 45° of grazing incidence and 23.8-29.5 nm at 50°) spectral range associated to a 0.1 nm step is used for the accurate measurement of the reflectance value while a wide (20.7-31.0 nm at both 45 and 50°) spectral range associated to a 0.2 nm step is considered for the estimation of the background.

## 3. Results and discussion

*3.1 X-ray reflectivity at 0.154 nm*

As an illustration, the x-ray reflectivity curves of the Mg/Co and Mg/Co/Zr multilayers measured at 0.154 nm are presented in Figure 1 on a logarithmic scale. Within the entire probed angular range, up to 11 well-defined Bragg peaks are present giving evidence for the good structural quality of the stacks. For Mg/Co, the seventh Bragg peak is almost extinguished as a consequence of the ratio of the Co thickness to the d period close to 1/7. It appears that, for all systems, the width of the measured Bragg peaks broadens as the grazing angle increases. This could be ascribed to a non homogeneity of the depth distribution of the layer thicknesses. The thickness, roughness and density values, deduced from the fit of the XRR curves, are collected in Table 2 where the density ratio (%) is defined as the density of the layer divided by the density of the bulk. The density values are given within a 5% uncertainty. For comparison, the value of the period is also calculated using the refraction corrected Bragg law.

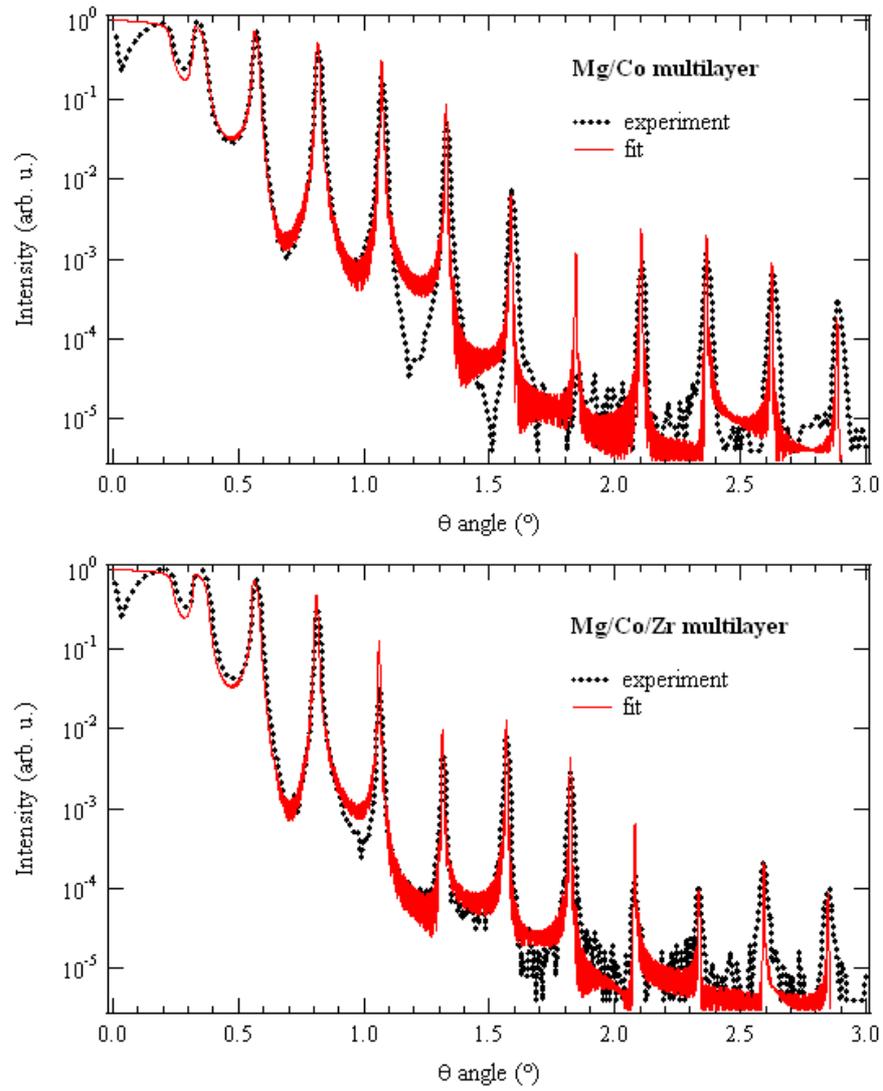

Figure 1: fit of the x-ray reflectivity curves of the Mg/Co and Mg/Co/Zr multilayers measured at 0.154 nm.

Table 2: structural parameters extracted from the fit of the XRR curves measured at 0.154 nm. For comparison, the period value is also calculated using the refraction corrected Bragg law.

| Multilayer | Bragg law corrected for refraction Period d (nm) | XRR fitting procedure | | | |
|---|---|---|---|---|---|
| | | d (nm) | $d_{co}$ (nm) $d_{Mg}$ (nm) $d_{Zr}$ (nm) | $\sigma$ (nm) | Density ratio Co (%) Density ratio Mg (%) Density ratio Zr (%) |
| Mg/Co | 16.88 | 16.9 | Co: 2.6 | $\sigma_{Co/Mg}$ = 0.5 | 100 |
| | | | Mg: 14.3 | $\sigma_{Mg/Co}$ = 0.6 | 96 |
| | | | - | | - |
| Mg/Zr/Co | 17.28 | 17.3 | Co: 2.3 | $\sigma_{Co/Zr}$ = 0.6 | 96 |
| | | | Mg: 13.5 | $\sigma_{Zr/Mg}$ = 0.6 | 100 |
| | | | Zr: 1.5 | $\sigma_{Mg/Co}$ = 0.7 | 94 |
| Mg/Co/Zr | 17.11 | 17.1 | Co: 2.4 | $\sigma_{Zr/Co}$ = 0.6 | 100 |
| | | | Mg: 13.2 | $\sigma_{Co/Mg}$ = 0.6 | 95 |
| | | | Zr: 1.5 | $\sigma_{Mg/Zr}$ = 0.7 | 98 |
| Mg/Zr/Co/Zr | 17.03 | 17.0 | Co: 2.0 | $\sigma_{Zr/Co}$ = 0.7 | 99 |
| | | | Zr: 1.5 | $\sigma_{Co/Zr}$ = 0.7 | 100 |
| | | | Mg: 12.0 | $\sigma_{Zr/Mg}$ = 0.5 | 100 |
| | | | Zr: 1.5 | $\sigma_{Mg/Zr}$ = 0.6 | 100 |

The measured periods are in good agreement with the aimed values while the roughness values are of the same magnitude for all structures. Within the estimated uncertainty, the layer densities are close to those of the bulk materials.

*3.2 EUV reflectivity*

The EUV reflectivity spectra of the Mg/Co, Mg/Zr/Co, Mg/Co/Zr and Mg/Zr/Co/Zr multilayers measured in the short spectral range at 45 and 50° of grazing incidence are presented in Figure 2. The asymmetric shape of the curves originates from the presence of the Mg L edge. For each multilayer, the Bragg peak stands onto a background corresponding to the total reflection and diffuse scattering. In the following, we will distinguish the values (in %) at the application wavelength of the peak reflectivity, the background and the net reflectivity, this latter being the difference between the two previous quantities. The background height at the application wavelength is estimated considering the wide (20.7-31.0 nm) spectral range.

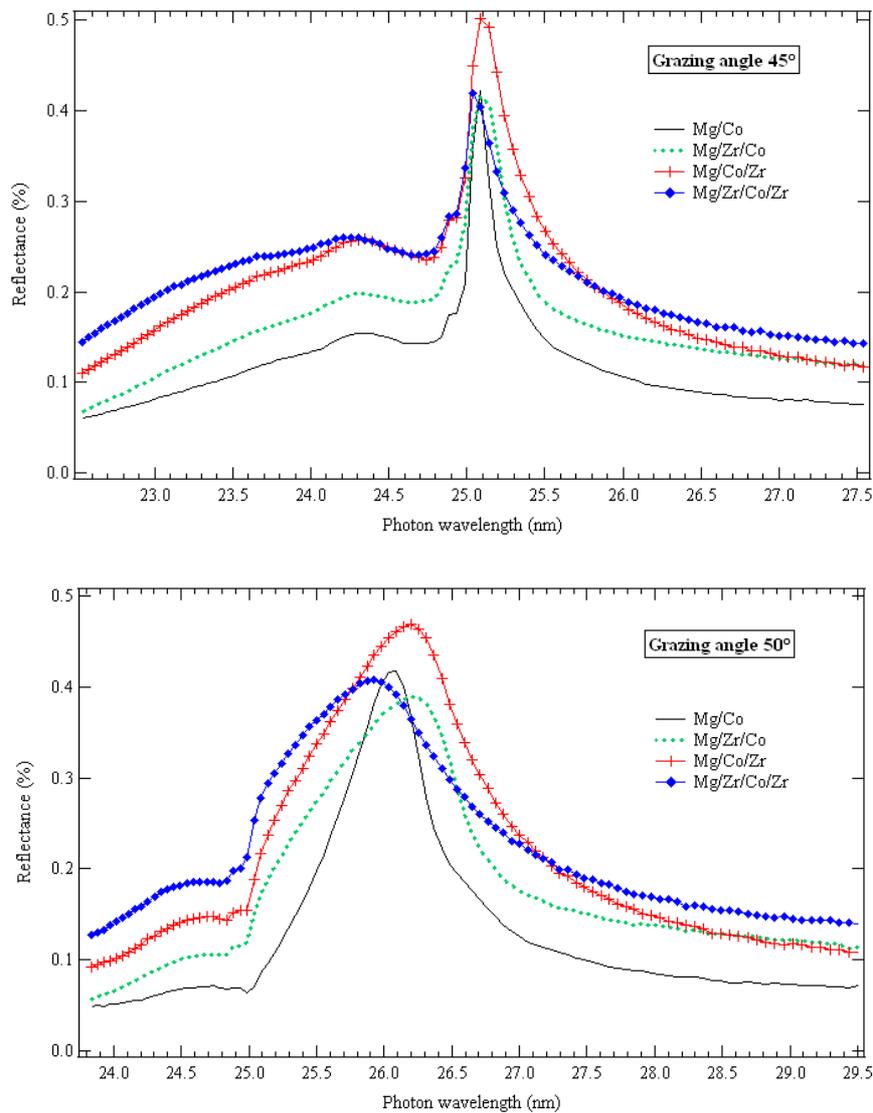

Figure 2: EUV reflectivity spectra of the Mg/Zr/Co, Mg/Co/Zr and Mg/Zr/Co/Zr multilayers measured at 45° (top) and 50° (bottom) of grazing incidence.

At 45°, the reflectivity peaks are narrow due to the absorption at wavelengths shorter than the Mg L edge. At 50°, the peak bandwidth is about twice larger than at 45°. The values of the peak reflectivity and background measured at 45° of grazing incidence in the wide spectral range (20.7-31.0 nm) are collected in Table 3. At 45°, with respect to the bi-layered Mg/Co structure (42.4%), the addition of a third component in the stack leads to a higher peak reflectivity (50.0%) if Zr is introduced at the Mg-on-Co interface while a lower value (41.4%) is measured when Zr is introduced at the Co-on-Mg interface. The presently studied quadri-layer leads to the lowest reflectance (40.6%) as expected from the IMD simulations. The background value is the highest for the quadri-layered system. Among the tri-layered structures, although Mg/Co/Zr exhibits the highest background, its net reflectivity remains the best. We do not present the peak reflectivity and background detailed values in the case of the 50° grazing incidence since these multilayers are designed for a 45° application angle.

Table 3: reflectivity and background values measured at 45° of grazing incidence in the wide spectral range (20.7-31.0 nm). $R_{exp}$ and $R_{sim}$ stand for measured and simulated reflectance respectively.

| Multilayer | $\lambda$ | Peak reflectivity | Background | Net reflectivity (%) | $R_{exp}/R_{simul}$ |
|---|---|---|---|---|---|
| Mg/Co | 25.1 | 42.4 | 5.5 | 36.9 | 0.75 |
| Mg/Zr/Co | 25.1 | 41.4 | 5.2 | 36.2 | 0.76 |
| Mg/Co/Zr | 25.1 | 50.0 | 6.4 | 43.6 | 0.80 |
| Mg/Zr/Co/Zr | 25.1 | 40.6 | 6.9 | 33.7 | 0.86 |

In spite of the relatively high background, since the reflectance at the wavelength of interest is enhanced with respect to the neighbouring wavelengths, these multilayers (particularly Mg/Co/Zr) could be efficiently used for imaging purposes in astrophysics applications for instance. As an illustration, Mg-based multilayers (Co/Mg, SiC/Mg, $B_4C$/Mg and Si/Mg) operated at near normal incidence have been very recently developed for solar He-II radiation at 30.4 nm [4]. On the contrary, their use for spectroscopic purpose may be inadequate since in that case high spectral purity is required.

To refine the description of each stack, the thickness, roughness and density values deduced from the fit of the XRR curves are introduced as constrained parameters to simulate the EUV reflectivity curves. The result is presented in Figure 3 for the Mg/Co, Mg/Zr/Co, Mg/Co/Zr and Mg/Zr/Co/Zr multilayers. It clearly appears that, for each multilayer, the simulated EUV peak reflectivity is systematically higher than the measured value. Three reasons can be proposed to explain this discrepancy:

- the optical constants of materials are obviously different at 0.154 nm and in the ~25 nm range. Moreover, the EUV refractive indices are not known with accuracy and the reflectivity strongly depends on these parameters;
- in XRR measurements, the grazing angle is in the range 0–3° whereas it is 45 or 50° for the EUV measurements. Then, the discrepancy could be due to the different penetration depth or to the different power spectral densities of the interface roughness of the layer surface;
- the presence of highly reactive materials, as for example Mg, could induce the formation of interface compounds whose optical properties are also not well known. Concerning this latter argument, we have recently demonstrated that in Co/Mg the interfaces are abrupt [3].

Nevertheless, in agreement with the XRR measurements at 0.154 nm, the structural quality of the stack is high since the ratio of the experimental to simulated reflectances is around 80%.

To improve again the agreement between the measured and simulated EUV reflectivity curves, we propose the following procedure based on the fact that, as mentioned above, in the EUV domain the optical indices (and material densities) are not known with a high accuracy. We consider the Co, Mg and Zr refractive indices ($n + ik$) as independent parameters to adjust in the EUV reflectivity simulations. With respect to the values collected in the CXRO database [http://www-cxro.lbl.gov], we consider a global variation (in %), independent increase or decrease of the real and imaginary parts.

The values of thickness and roughness of each layer are kept constant to the values determined through the XRR fit (Table 2).

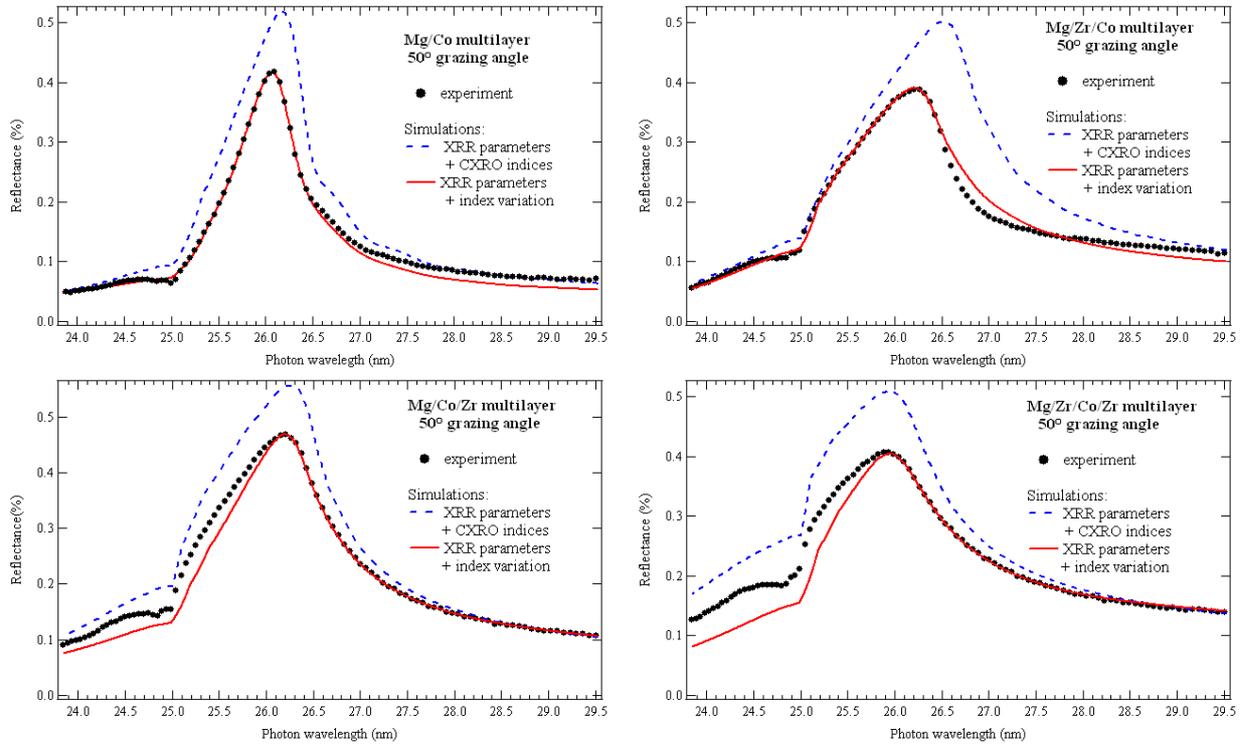

Figure 3: Measured (dotted line) EUV reflectivity curves compared to simulations using the thickness and roughness from the XRR fit and the CXRO optical indices (dashed line) and simulations using the thickness and roughness of the XRR fit and the variation of the optical indices (solid line).

Table 4: variations, with respect to the CXRO database, of the optical complex indices $n + \mathrm{i}k$ of Co, Mg and Zr in order to fit the EUV reflectivity curves.

| Multilayer | $n$ index variation (%)<br>Co<br>Mg<br>Zr | $k$ index variation (%)<br>Co<br>Mg<br>Zr |
|---|---|---|
| Mg/Co | 107<br>99<br>- | 115<br>115<br>- |
| Mg/Zr/Co | 107<br>99<br>83 | 115<br>115<br>100 |
| Mg/Co/Zr | 107<br>99<br>108 | 115<br>115<br>140 |
| Mg/Zr(1)/Co/Zr(2) | 107<br>99<br>Zr(1) 112<br>Zr(2) 108 | 115<br>115<br>110<br>140 |

First, in the Mg/Co multilayer, with respect to the CXRO data, the real and imaginary parts have to be increased for Co while for Mg only the imaginary part has to be enhanced keeping almost constant the real part. Then, in the tri- and quadri-layered structures, the variations found for Co and Mg in Mg/Co are frozen and only the Zr indices are adjusted. The need for large variations reveals the lack of precise knowledge of the indices in the EUV range.

*3.3 Comparison with the theoretical predictions*

For tri- and quadri-layered structures, we consider the predictions of the theoretical model developed by Larruquert and devoted to the EUV reflectance enhancement in multilayers with more than two materials [8, 11, 14]. To ensure the highest reflectance, the model considers the largest possible refractive index contrast among the materials and ends up in a material selection rule within a period. Nevertheless, in his model, Larruquert has excluded the highly reactive materials such as alkaline and most alkalino-earth metals (Mg). In addition, given the Mg layer thickness (12 to 14.45 nm), we have to note that the presently studied multilayers are not sub-quarterwave multilayers.

Within the multilayer, we consider a period made of $m$ thin films of different absorbing materials deposited onto the substrate and the outer medium is designated using the "inc" subscript. The letter $i$ is chosen to identify a given layer within the stack: $i = 1$ corresponds to the outermost layer and $i = m + 1$ the substrate. For the tri-layered structure, $m$ is equal to 3. For $s$-polarized incoming radiation, two conditions are required to lead to maximum reflectance [11]:

$$\begin{cases} \text{Im}\,[(s\Delta N_1)/\cos\theta_1] > 0 & \text{condition 1} \\ \text{Im}\,[\Delta N_i/\Delta N_{i-1}] < 0 \quad i = 2, \ldots, m & \text{condition 2} \end{cases}$$

where $s = r^*_{inc,1}(1 - r^2_{inc,1})$ in which $r_{inc,1}$ stands for the amplitude reflectance given by the Fresnel coefficient at the interface with refractive indices $N_{inc}$ and $N_1$ and $\Delta N_i = N_{i+1} - N_i$. In summary, to ensure an enhanced reflectivity, the condition 1 should lead to a positive value and the condition 2 to a negative value. Table 5 collects the values resulting from the calculation of conditions 1 and 2 in the case of the Mg, Co and Zr materials. We have successively considered 3 material sequences, namely Zr/Mg/Co, Mg/Zr/Co and Mg/Co/Zr where the first layer on the substrate is the first material given in the sample name. The Mg/Zr/Co/Zr quadri-layered structure is not taken into account since it is made of only three different materials.

Table 5: application of conditions 1 and 2 developed by Larruquert [11] in the case of the Mg, Co and Zr materials. Optical constants originate from the CXRO database [http://www-cxro.lbl.gov].

| Multilayer | Condition 1 | Condition 2 | |
|---|---|---|---|
| | - | $i = 2$ | $i = 3$ |
| Substrate/Zr/Mg/Co | + 0.011 | - 0.221 | - 0,090 |
| Substrate/Mg/Zr/Co | - 0.016 | + 0.623 | + 0,090 |
| Substrate/Mg/Co/Zr | + 0.029 | - 0.623 | - 0,221 |

From these results, the model predicts that the Mg/Co/Zr structure is the most suited in terms of enhanced EUV reflectivity. Indeed, it is the only structure fulfilling the two conditions.

An alternative way of using the theoretical model of Larruquert is to adopt a graphical method [8]. In Figure 4, we have plotted in the *n-k* plane the optical constants at $\lambda = 26.0$ nm of a large set of materials where the refractive index $N = n + ik$. Data are extracted from the CXRO database. According to Reference [8], the "correct" material order leading to an enhanced reflectivity is as follows: clockwise rotation corresponds to the top layer to substrate sequence.

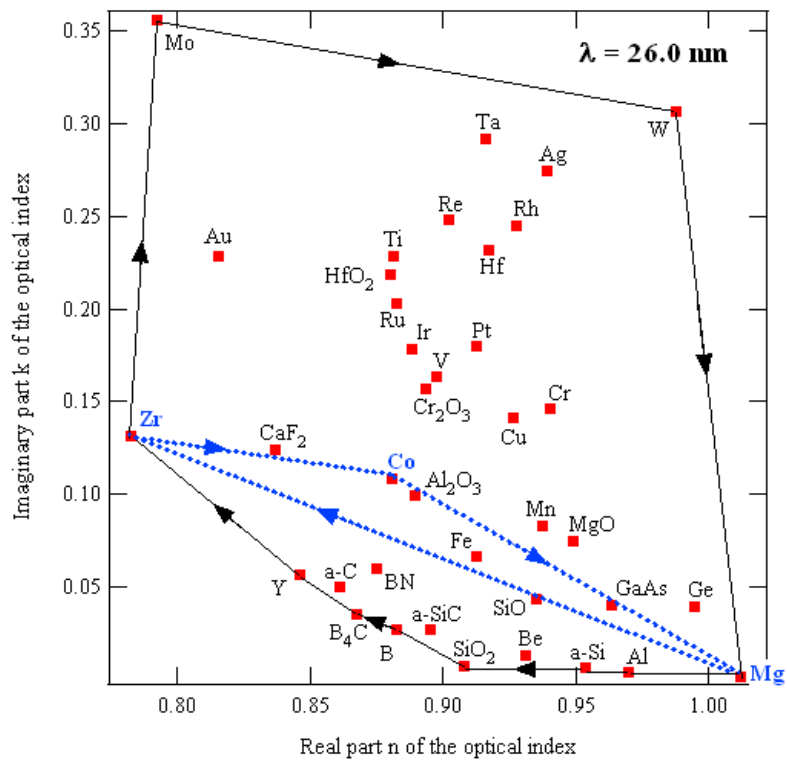

Figure 4: determination of the material order in the tri-layered structure applying the model of Larruquert [8] at $\lambda = 26.0$ nm.

The so-called "minimum polygon" including all the materials, namely Zr/Mo/W/Mg/Al/SiO$_2$/B$_4$C/Y, is plotted in solid line and presents the largest circumference. The multilayer with an optimized film thickness distribution including all the materials of the minimum polygon will lead to a reflectance higher than the reflectance that can be obtained with a multilayer constructed with a subset of these materials. Nevertheless, an important assumption is considered: the interfaces are assumed to be smooth and abrupt, with no interdiffusion of the materials across the interface. The selection of a set of materials whose optical constants are inside the minimum polygon will lead to a lower reflectance regardless of the number of materials. As a consequence, in our case, the secondary polygon (dotted line), that is to say $Zr/Co/Mg/Al/SiO_2/B_4C/Y$, will be characterized by a lower reflectivity than that for the "minimum polygon" since its circumference is smaller. It appears that the substrate/Mg/Co/Zr sequence is more suited in terms of reflectivity than the substrate/Mg/Zr/Co combination.

From the secondary polygon, the $Mg/Co/B_4C$ structure should also be pertinent according to the model. Nevertheless, in our recent study on $Co/Mg/B_4C$ multilayers, nuclear magnetic resonance (NMR) spectroscopy has allow us to evidence a strong intermixing between Co atoms and B and/or C atoms from $B_4C$ layers leading to a reflectivity of 0.7% at 25.1 nm [3].

*3.4 Thermal stability of the Mg/Co and Mg/Co/Zr multilayers*

We have studied the thermal stability of the Mg/Co and Mg/Co/Zr multilayers up to an annealing temperature equal to 200°C. The evolution, as a function of the annealing temperature, of the reflectivity of the Mg/Co and Mg/Co/Zr multilayers measured in the short spectral range at 45 and 50° of grazing incidence is presented in Figure 5. For both angles (even if this effect is more pronounced at 50°), as the annealing temperature increases, the Bragg peak position is slightly shifted towards higher wavelengths. This shift away from the Mg L edge may have for consequence the slight reflectivity increase observed for both samples at 50°. Since the peak position is related to the period of the stack, the peak shift towards higher wavelengths would correspond to a slight period increase (period dilatation) with the annealing temperature. Since we have not performed XRR measurements on the annealed multilayers, we cannot follow the evolution of the period value as a function of the annealing temperature. The values of the peak reflectivity and background measured at 45° of grazing incidence in the wide spectral range (20.7-31.0 nm) are presented in Table 6 as a function of the annealing temperature.

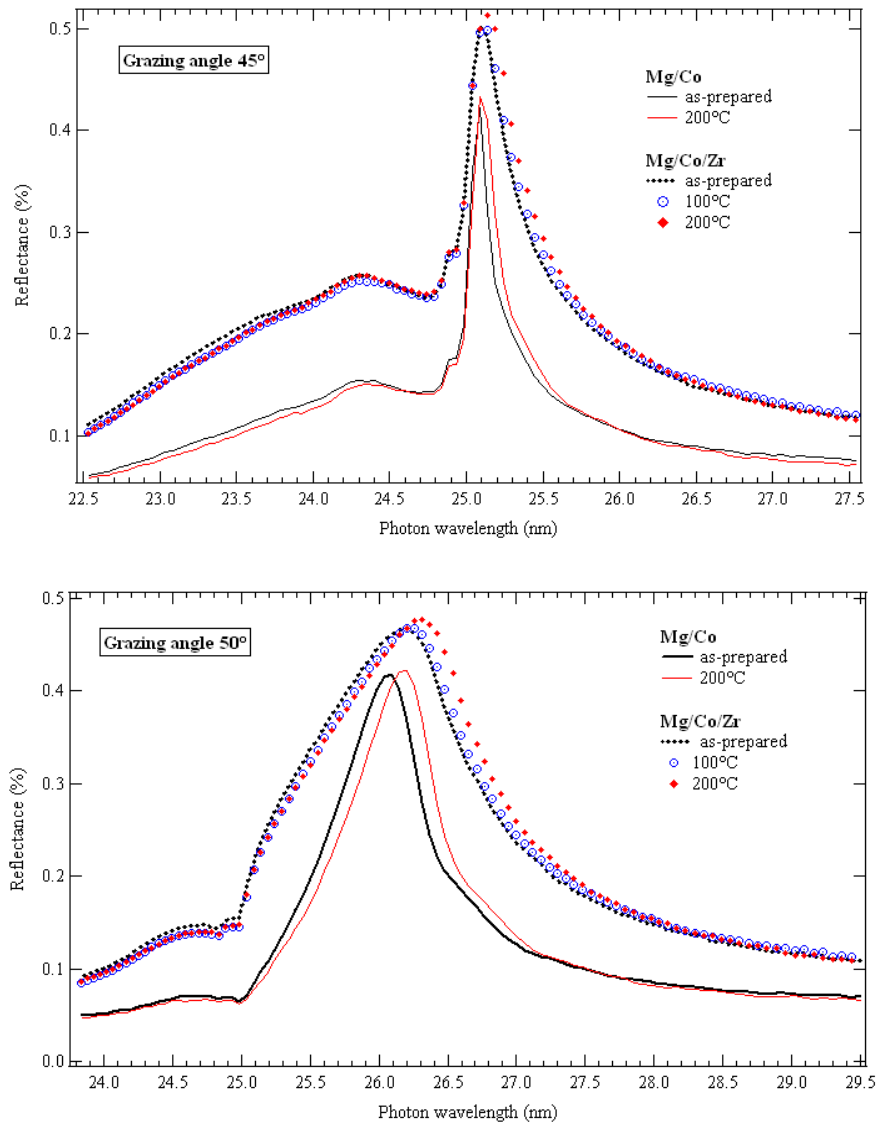

Figure 5: Evolution, as a function of the annealing temperature, of the Mg/Co and Mg/Co/Zr EUV reflectivity spectrum measured at 45 and 50° of grazing incidence in the short spectral range.

Table 6: Evolution as a function of the annealing temperature of the values, at 45° of grazing incidence, of the peak reflectivity and background measured in the wide spectral range (20.7-31.0 nm).

| Multilayer | Mg/Co | | Mg/Co/Zr | |
|---|---|---|---|---|
| | Peak reflectivity | Background | Peak reflectivity | Background |
| As-prepared | 42.4% @ 25.1 nm | 6.0% @ 25.1 nm | 50.0% @ 25.1 nm | 6.9% @ 25.1 nm |
| 100°C | - | - | 49.6% @ 25.1 nm | 6.8% @ 25.1 nm |
| 200°C | 44.1% @ 25.1 nm | 5.6% @ 25.1 nm | 51.3% @ 25.1 nm | 7.0% @ 25.1 nm |

For both the bi- and tri-layered structures, the reflectivity slightly increases with the temperature. At 45° of grazing incidence and considering the samples annealed at 200°C, the peak reflectivity value is at 25.1 nm equal to 44.1% for Mg/Co and 51.3% for Mg/Co/Zr. This latter value constitutes a promising result as far as the parameters during the multilayer deposition can be optimized.

## 4. Conclusion

We have presented the comparative analysis of the EUV optical quality of bi-, tri- and quadri-layered structures based on Mg, Co and Zr. The stack structural parameters and EUV optical performances have been estimated through X-ray and EUV reflectometry measurements respectively. The addition of Zr at the only Mg-on-Co interface has proven to be an efficient combination to enhance the reflectance. The analysis of the measured EUV reflectance of tri- and quadri-layered systems has allowed us to validate the model of Larruquert predicting the order of the materials introduced within the structure. A reflectance slightly higher than 50% at 25.1 nm is reported for the Mg/Co/Zr system annealed at 200°C. Nevertheless it would be pertinent to perform complementary analysis, such as X-ray reflectometry, to investigate the slight period dilatation upon annealing. Given the compromise between the high reflectance at the application wavelength and the relatively high background, stress is laid on the use of these multilayers for imaging purpose through astrophysics applications for example.